\documentclass[12pt]{article}
\usepackage[dvips]{graphicx}
\usepackage{color}
\usepackage{amsfonts,amssymb}
\begin{document}

\vspace*{1cm}
\begin{center}
{\Large \bf The Nucleon and Delta Baryons\\[1ex]
in the Unified Picture for Hadron Spectra}\\

\vspace{4mm}

{\large A. A. Arkhipov\footnote{E-mail: Andrei.Arkhipov@ihep.ru}\\
{\it State Research Center ``Institute for High Energy Physics" \\
 142281 Protvino, Moscow Region, Russia}}\\
\end{center}

\vspace{2mm}
\begin{abstract}
In this article it has been shown that the recent PDG Baryon Particle
Listings of the Nucleon and Delta states, including some evidence for
new states, have excellently incorporated in the unified picture for
hadron spectra created earlier \cite{8}. It is claimed that this is a
strong confirmation of our theoretical concept. A comparison with
experiment is briefly discussed.
\end{abstract}

\section{Introduction}

The physics of nucleon resonances has always been a hot matter of
discussion in hadronic spectroscopy. It is well known that
conventional scheme of classification, systematics and interpretation
of all hadronic states \cite{1} is based on the constituent quark
model in its many different versions. In the simplest variant of the
constituent quark model constructed in the early years all known
lightest hadrons made of just three $u$, $d$ and $s$ quarks were
classified in according to irreducible representations of the
$SU(3)_f$ group where mesons were made out of $q\bar q$, while
baryons were built from $qqq$. By this way the lowest $q\bar q$ meson
configuration can be decomposed as $\bf3\otimes\bar{\bf 3}=\bf
8\oplus\bf1$, while the lowest $qqq$ baryon configuration can be
decomposed $\bf3\otimes\bf3\otimes\bf3=\bf 10\oplus\bf 8\oplus\bf
8\oplus\bf 1$ as well. It is a remarkable fact that octet
($\pi^0,\pi^+,\pi^-,K^0,K^+,\bar{K^0},K^-,\eta$) of mesons and octet
($p,n,\Sigma^0,\Sigma^+,\Sigma^-,\Xi^0,\Xi^-,\Lambda$) and decuplet
($\Delta^0,\Delta^-,\Delta^+,\Delta^{++},
\Sigma^{*0},\Sigma^{*+},\Sigma^{*-},\Xi^{*0},\Xi^{*-},\Omega^-$) of
baryons have experimentally been observed. The experimental discovery
of the predicted $\Omega^-$ hyperon was a shining confirmation of
$SU(3)_f$ symmetry and of its important role in the classification
and systematics of hadronic states. The addition of the $c$, $b$, and
$t$ quarks to the above three light quarks extends, in principle, the
flavor symmetry to $SU(6)_f$. However, earlier the $SU(6)$ group has
been considered as an approximate spin-flavor symmetry for the
baryons made from just $u$, $d$ and $s$ quarks (see e.g. \cite{2}).
In that case the baryons are classified by the multiplets arising in
the decomposition
$\bf6\otimes\bf6\otimes\bf6=\bf{56}\oplus\bf{70}\oplus\bf{70}\oplus\bf{20}$.
Here, the basic states are
$u_{\uparrow},u_{\downarrow},d_{\uparrow},d_{\downarrow},s_{\uparrow},
s_{\downarrow}$ where $\uparrow$ and $\downarrow$ denote spin up and
down. Next, the $SU(6)$ multiplets decompose into $SU(3)_f$
multiplets ${\bf56}={}^4{\bf10}\oplus{}^2\bf8$,
${\bf70}={}^2{\bf10}\oplus{}^4{\bf8}\oplus{}^2{\bf8}\oplus{}^2\bf1$,
${\bf20}={}^2{\bf8}\oplus{}^4\bf1$, where the superscript $(2S+1)$
represents the total spin $S$ of the quarks for all particles in the
given $SU(3)_f$ multiplet. So, the above mentioned baryon's octet
containing the nucleon, and the decuplet containing $\Delta$(1232)
belong to one and the same $SU(6)$ multiplet (56-plet ) which might
be considered as a lowest state where the orbital angular momenta of
all quark pairs are zero. Then the $\bf70$ and $\bf20$ could refer to
the states with nonzero orbital angular momenta of quark pairs or
something else. In this case the states with nonzero orbital angular
momenta may be classified by $SU(6)\otimes O(3)$ supermultiplets.
Even though the $SU(6)$ symmetry is broken by spin dependent
interactions the $SU(6)\otimes O(3)$ basis was a suitable one for
representing the baryon states. However, here the problem of so
called ``missing''  quark-model states arises, it has no solution so
far. Of course, in that case one could imagine some selection rules
which are responsible for the fact that many baryons have not been
observed. At the same time, many recent experiments have reported the
observation narrow structures which cannot be explained by the
standard quark-model assignments for baryons as $qqq$ states. This,
first of all, concerns the number of narrow baryon structures
observed in the missing mass $M_X$ and in the $p\pi^+$ invariant mass
distribution in the reaction $pp\rightarrow p\pi^+X$, which cannot be
associated with the standard $qqq$ quark configurations
\cite{3,4}.\footnote{I thank B.~Tatischeff  for drawing my attention
to Ref. 3 caused this study.} Why these baryon states are less
massive and so narrow than predicted in quark models is an open
question so far. Certainly, this raises a challenge to the theory.
The other non-$qqq$ baryons have been observed as sharp structures in
the $nK^+$ and $pK^0$ invariant mass distribution, now denoted as the
$\Theta^+$ baryons \cite{5}, and in the $\Xi^{-} \pi^\pm$,
$\bar\Xi^{+}\pi^\pm$ invariant mass spectra \cite{6} as well -- all
of that was interpreted as candidates for pentaquarks ($qqqq\bar q$).
Here, it should also be mentioned an experimentally well established
evidence for many  non-$q\bar q$ meson's states (see e.g. discussion
in Ref. \cite{10} and references therein) declared often and often as
exotics. In fact, exotic simply includes all hadrons which cannot be
explained in the framework of the simple valence picture of $q\bar q$
for mesons or $qqq$ for baryons. Note, that even though the simple
valence picture operates degrees of freedom like QCD fields the
valence quarks are not identical to QCD fields.

The recent PDG Baryon Particle Listings contain 22 Nucleon and 22
Delta states which are the excited states of the nucleon observed in
a large number of formation and production experiments. The
conventional masses, widths and other discrete quantum numbers of the
$N$ and $\Delta$ resonances in the PDG Baryon Listings have largely
been defined from partial-wave analysis of $\pi N$ scattering data.
However, any specific constituent quark models even though with a
clear set of dynamical ingredients provide quite an another
assignments for the quantum numbers of baryons as $qqq$ states and
predict a much richer spectrum of baryon states than has been found
in partial-wave analysis of $\pi N$ scattering data. That is why,
many attempts have been undertaken to search for ``missing'' or
``hidden'' quark-model states from partial-wave analysis in the
production processes of other final states such as $N2\pi$, $N\rho$,
$N\eta$, $N\omega$, $\Lambda K$, $\Sigma K$ (see e.g. \cite{7}).
Besides, there is a serious problem to translate the results of a
constituent quark model calculations and predictions into the
standard partial-wave analysis conventions accepted by PDG. That
translation cannot be constructed in the framework of a given
constituent quark model without some additional assumptions and
conventions. At any rate, to make such translation in a more clear
way we have to consider pentaquark states even for the usual,
non-exotic $N$ and $\Delta$ baryons. The consistent dynamical
description of pentaquark states is very tedious and hard labour
which has not been done yet, if not impossible at all. That sheer
drudgery will unlikely done in the near future because the consistent
dynamical description of more simple three-quark states though has
not been performed so far.

Concerning the dynamical content of constituent quark models one
could say that any specific quark model even with some
``QCD-inspired'' improvements is a phenomenological, non-relativistic
potential model without a reliable ground in quantum field theory, in
particular, in QCD. For instance, the usual used constituent quark
mass parameters of about 300 MeV for the light $u$ and $d$ quarks
cannot be derived in the framework of the underlying QCD.  Of course
quarks can move relativistically and this also means that theoretical
consistency of a non-relativistic potential model is most likely
doubtful. We would like also to point out that constituent quark
models with the current quark masses about a few MeV have a serious
problem with the value of the nucleon sigma term measured in low
energy $\pi N$ scattering. To resolve the $\sigma$-term problem the
strong Chiral Symmetry Breaking is needed, and the mechanism for that
is not clear so far. Nevertheless, there is understanding that such
mechanism might be found in the framework of non-perturbative quantum
field theory.

Moreover, there is a hope that lattice computations in QCD with a
powerful computers could help us to eliminate all imperfections of
non-relativistic quark potential models, if lattice studies can help
us at all. Time will show.

Recently a new, very simple and quite general theoretical concept
concerning the structure of hadron spectra has been formulated which
allowed to construct the global solution of the spectral problem in
hadron spectroscopy; see \cite{8} and references therein where some
of our previous studies were partially summarized. It has been
claimed that existence of the extra dimensions in the spirit of
Kaluza and Klein together with some novel dynamical ideas may provide
new conceptual issues and quite new scheme of systematics for hadron
states to build the unified picture for hadron spectra up. The main
advantage of the developed theoretical concept is that all calculated
numbers for masses and widths of hadron states do not depend on a
special dynamical model but follow from fundamental hypothesis on
existence of the extra dimensions with a compact internal extra
space. One very important fact has been established in a reliable
way: the size of the internal compact extra space determines the
global characteristics of the hadron spectra while the masses of the
decay products are the fundamental parameters of the compound systems
being the elements of the global structure. What is remarkable that
all new hadron states experimentally discovered last years have been
observed just at the masses predicted in our approach, and those
states appeared to be narrow as predicted too. A thorough analysis
\cite{9} of many different experiments reported the observation of a
new very narrow, manifestly exotic $\Theta^+$ (Q=1, S=1) baryon, with
the simplest quark assignment ($uudd\bar s$) decaying into $nK^+$ and
$pK_S^0$, taken together allowed us to claim that many different
$\Theta$ states have been discovered and all of them were excellently
incorporated in the unified picture for hadron spectra developed.
This concerns the newly discovered $\Xi_{3/2}^{--}$ baryon with
strangeness  $S =-2$, isospin I = $\frac{3}{2}$ and a quark content
of ($dsds\bar{u}$) \cite{6}, now denoted as $\Phi^{--}$ by PDG, as
well. We have also shown that a large amount of experimental data may
excellently be incorporated in the systematics provided by the
created unified picture for hadron spectra. In this article we apply
our approach to show what place in the unified picture for hadron
spectra the $N$ and $\Delta$ baryons take up.

\section{Understanding the $N$ and  $\Delta$ baryons \\
in the unified picture for hadron spectra}

According to the general, theoretical concept \cite{8} we calculate
the Kaluza-Klein tower of KK-excitations for the $N\pi$ system by the
formula

\begin{equation}\label{Npi}
M_n^{N\pi} = \sqrt{m_{N}^2+\frac{n^2}{R^2}} +
\sqrt{m_{\pi}^2+\frac{n^2}{R^2}}\,,\quad (n=1,2,3,...),
\end{equation}
where $R$ is the same fundamental scale established before (see
\cite{8} and references therein for the details), $N=(p,n)$,
$\pi=(\pi^0, \pi^\pm)$, and the masses of proton, neutron and pions
have been taken from PDG. The such built Kaluza-Klein tower is shown
in Tables 1--5. For simplicity we have considered one-dimensional
compact internal extra space and only diagonal elements of the
Kaluza-Klein tower have been presented. The experimental data
extracted from PDG \cite{1} and Refs. \cite{3,4} are shown in Tables
1--5 as well. The data from Refs. \cite{3,4} only are shown in
separate Table 5. As seen all narrow low mass baryons shown in the
Tables are in excellent agreement with the calculated KK excitations.
The other narrow low mass baryon's structures found in Refs.
\cite{3,4}, if any,  might be identified with non-diagonal elements
of the KK tower for the $N\pi$ system. As a rule, non-diagonal
elements of the KK towers are suppressed in reality; see however
\cite{8} for the details. We would like to emphasize that the
so-called universal internal toroidal extra spaces might be
considered as a natural explanation of suppression for non-diagonal
elements of the KK towers by conservation low of KK numbers. In other
words, an experimental observation of hadronic states corresponding
to non-diagonal elements of the KK towers could be considered as an
evidence for existence of generic internal compact extra spaces.

Our conservative estimate for the widths of KK excitations looks like
\begin{equation}\label{width}
\Gamma_n \sim \frac{\alpha}{2}\cdot\frac{n}{R}\sim 0.4\cdot n\,
\mbox{MeV},
\end{equation}
where $n$ is the number of KK excitation, and $\alpha \sim 0.02$,
$R^{-1}=41.48\,\mbox{MeV}$ are known from our previous studies
\cite{8}. This model independent estimate is universal for all of the
KK towers, it does not depend on a composition of the compound
systems living there. Certainly, some model dependent dynamics might
modify this estimate. However, one property of estimate (\ref{width})
is an absolute evidence for the higher the KK excitation is, the
larger is the width of the corresponding compound system. This
property  has to be fulfilled in any model. The broad peaks in the
hadron spectra are interpreted in our approach as an envelope of the
narrow KK excitations predicted by the Kaluza-Klein scenario.

The most of the nucleon resonances presented in the PDG Baryon
Particle Listings have been extracted from partial wave analysis
performed by a few groups: the Carnegie-Mellon Berkeley (CMB) group
\cite{11}, the Karlsruhe-Helsinki (KH) group \cite{12} and the Kent
State University (KSU) group \cite{13} are the most famous among of
them. It should also be noted the article \cite{14} where the CMB
analysis has significantly been extended with account of a larger
data set including the modern data at the moment. In fact, the
formalism in Ref. \cite{14} is identical to CMB but the data base
used is similar to KSU. We would also like to mention the old paper
\cite{15} and review article \cite{16}. Each group has an own
``cookery'' in preparing the results of the analysis, the performed
analyses differ from each other often significantly in the methods
and the data sets used to extract the resonances, that is why, there
exist sometimes large enough discrepancies between different groups.

Tables 6-7 compare the results of the KSU, KH and CMB analyses with
the values of KK excitations for the $p\pi$ system. The best
agreements with the KK excitations values are shown in Tables 6-7 by
the bold-face numbers.

We have also calculated the Kaluza-Klein towers of KK-excitations for
the $N\rho$, $N\eta$, $N\omega$, $\Lambda K$, $\Sigma K$ systems by
the formulae similar to (\ref{Npi}), and for the $N2\pi$ system by
the formula
\begin{equation}\label{N2pi}
M_n^{N2\pi} = \sqrt{m_{N}^2+\frac{n^2}{R^2}} +
\sqrt{m_{\pi^1}^2+\frac{n^2}{R^2}} +
\sqrt{m_{\pi^2}^2+\frac{n^2}{R^2}} \,,\quad (n=1,2,3,...),
\end{equation}
as this has been prescribed in \cite{8}; here, as above, $N=(p,n)$,
$\pi^1(\pi^2)=(\pi^0, \pi^\pm)$. These KK towers are shown in Tables
8-15. As above, we have restricted ourselves by the simplest case of
one-dimensional compact internal extra space and only diagonal
elements of the Kaluza-Klein towers have been presented. The
arrangement of the known Nucleon and Delta baryons in the
Kaluza-Klein towers is presented in Tables 8-15\footnote{Even though
the $\Delta$ states in the $N\eta$, $N\omega$, and $\Lambda K$
systems are forbidden by isospin we have listed the $\Delta$ baryons
in Tables 11-13 for convenience too.} as well.

\section{Discussion of comparison with experiment}

As seen from the Tables all experimentally observed Nucleon and Delta
baryons including narrow low mass baryons are excellently
accommodated within them. We see only one empty cell in the $N\pi$ KK
tower corresponding to the $M_{26}^{N\pi}$(1517) storey (see Tables
1-4), but very probably that this fact relates to our incomplete
knowledge of modern experimental data base. Sometimes one and the
same storey in the $N\pi$ KK tower is occupied by several baryons
with approximately equal masses within errors.

It should be emphasized one remarkable fact: we did not find a place
for the $P_{33}$(1232) baryon in Tables 3-4, the
$M_{5}^{N\pi}$(1208-1212) and $M_{6}^{N\pi}$(1254-1257) storeys are
not comfortable for this state. However, we found that the first
$M_{1}^{N2\pi}$(1222-1232) storey in the $N2\pi$ KK tower (see Tables
8-9) is just the place for the $P_{33}$(1232) baryon. This means that
the $P_{33}$(1232) baryon may have the true three-body origin;
really, the symbol $\Delta$ is quite appropriate one to correspond to
this fact. The other possibility is to search for the $P_{33}$(1232)
baryon among the non-diagonal elements in the $N\pi$ KK tower. For
example,
\[
M_{nm}^{p\pi^\pm}(n=3,m=6) = 1231.84\,\mbox{MeV},\quad
M_{nm}^{p\pi^\pm}(n=7,m=5) = 1232.17\,\mbox{MeV},
\]
\[
M_{nm}^{p\pi^\pm}(n=11,m=3) = 1230.33\,\mbox{MeV},\quad
M_{nm}^{n\pi^\pm}(n=11,m=3) = 1231.5\,\mbox{MeV},
\]
\[
M_{nm}^{n\pi^0}(n=3,m=6) = 1230.9\,\mbox{MeV},\quad
M_{nm}^{n\pi^0}(n=7,m=5) = 1230.87\,\mbox{MeV},
\]
where
\begin{equation}\label{Npitot}
M_{nm}^{N\pi} = \sqrt{m_{N}^2+\frac{n^2}{R^2}} +
\sqrt{m_{\pi}^2+\frac{m^2}{R^2}}\,,\quad (n,m=1,2,3,...).
\end{equation}
Tables 8-9 contain the arrangement of the other Delta baryons in the
$N2\pi$ KK tower too.

Obviously, that is noteworthy fact, which we would like to point out
here, concerning the $P_{11}$ resonance at 1462$\pm$10 MeV extracted
in Ref. \cite{13}. This resonance just occupies the
$M_{10}^{N\pi}$-storey in the $N\pi$ KK tower; see Table 1. Note,
that the quark-model calculations for the mass of this resonance give
1405 MeV and 1383 MeV \cite{13} that is in strong disagreement.

The new $P_{31}$ resonance at 1744$\pm$36 MeV found in \cite{13} is
also incorporated in our approach; see $M_{15}^{N\pi}$-storey in the
$N\pi$ KK tower in Table 2. The quark-model calculations for the mass
of the $P_{31}$ resonance give 1875 MeV and 1906 MeV \cite{13} which
are also in strong disagreement. The new $F_{35}$ resonance at
1752$\pm$32 MeV found in \cite{13} may occupy the same
$M_{15}^{N\pi}$-storey in the $N\pi$ KK tower.

The third $D_{13}$ resonance at 1804$\pm$55 MeV found in \cite{13}
has also an own place in the $N\pi$ KK tower and especially in the
$N\rho$ and $N\eta$ KK towers; see Tables 10-11. Here, one of the
quark-model predictions 1809 MeV for the masses of the $D_{13}$
resonances is in good agreement \cite{13}.

The second $P_{13}$ resonance at 1879$\pm$17 MeV, the third $P_{11}$
resonance at 1885$\pm$30 MeV, and the second $F_{15}$ resonance at
1903$\pm$87 MeV found in \cite{13} live on one and the same
$M_{17}^{N\pi}$-storey in the $N\pi$ KK tower; see Table 1. These
resonances have also comfortable places in the $N\rho$, $N\eta$,
$N\omega$, $\Lambda K$ and  $\Sigma K$ KK towers; see Tables 10-15.

Probably the $M_{18}^{N\pi}$-storey in the $N\pi$ KK tower is not so
good place for the $S_{31}$ resonance at 1920$\pm$24 MeV, and for the
$S_{11}$ resonance at 1928$\pm$59 MeV found in \cite{13}. However,
the $M_{12}^{\Lambda K}$-storey in the $\Lambda K$ KK tower (Table
13) is quite suitable for these resonances.

We would like to especially emphasize that the lowest $D_{35}$
resonance at 1956$\pm$22 MeV, the third $P_{33}$ resonance at
2014$\pm$16 MeV, the second $D_{33}$ resonance at 2057$\pm$110 MeV,
the lowest $F_{17}$ resonance at 2086$\pm$28 MeV, and the high-mass
$D_{35}$ resonance at 2171$\pm$18 MeV found in \cite{13} are
excellently incorporated in our approach; see Tables 1-4,10-15. The
masses for all of these resonances do not correspond to the
quark-model calculations \cite{13}.

It seems the $M_{21}^{N\pi}$-storey in the $N\pi$ KK tower is not so
comfortable for the lowest $G_{17}$ resonance at 2127$\pm$9 MeV found
in \cite{13}. Very probably that this resonance lives together with
the N[2100]$P_{11}$ and with the $F_{17}$ resonance at 2086$\pm$28
MeV on the same storey, as it's clear from Tables 1-4,10-13. In
addition, the same $G_{17}$ resonance at 2168$\pm$18 MeV found in
\cite{14} excellently corresponds to the $M_{21}^{N\pi}$-storey in
the $N\pi$ KK tower, and the $S_{31}$ resonance at 1802$\pm$87 MeV
extracted in \cite{14} is excellently incorporated in our approach
too; see Tables 1-4,10-15. .

We would like to mention too the $S_{11}$(1535) resonance at
1542$\pm$3 MeV (in accordance with $M_{5}^{N\eta}$; see detail
discussion in \cite{14}), the $P_{11}$(1440) resonance at 1479$\pm$80
MeV (in accordance with $M_{10}^{N\pi}$), the $P_{11}$(1710)
resonance at 1699$\pm$65 MeV (in accordance with $M_{14}^{N\pi}$),
the $P_{11}$(2100) resonance at 2084$\pm$93 MeV (in accordance with
$M_{20}^{N\pi}$), the $D_{13}$(1520) resonance at 1518$\pm$3 MeV (in
accordance with $M_{11}^{N\pi}$), the $G_{17}$(2190) resonance at
2168$\pm$18 MeV (in accordance with $M_{21}^{N\pi}$), the
$S_{31}$(1620) resonance at 1617$\pm$15 MeV (in accordance with
$M_{13}^{N\pi}$), the $S_{31}$(1900) resonance at 1802$\pm$87 MeV (in
accordance with $M_{16}^{N\pi}$), the $P_{31}$(1750) resonance at
1721$\pm$61 MeV (in accordance with $M_{15}^{N\pi}$), and the
$F_{35}$(1905) resonance at 1873$\pm$77 MeV (in accordance with
$M_{17}^{N\pi}$) found all in Ref. \cite{14} which are also
excellently incorporated in the unified picture for hadron spectra.

\section{Summary and conclusion}

This work should be considered as a continuation of our previous
studies concerning the structure of hadron spectra. We have
established that the recent PDG Baryon Particle Listings of Nucleon
and Delta states, including some evidence for new states, have
excellently incorporated in the theoretical concept developed earlier
\cite{8}. In particular, it was shown that new resonances found in
Ref. \cite{13}, including the $P_{31}$ state at 1744$\pm$36, the
$F_{35}$ state at 1752$\pm$32, and $P_{13}$ state at 1879$\pm$17
which did not predicted in the quark-model calculations, have
excellently accommodated in the corresponding KK towers. Moreover,
the recently reported narrow low mass baryons \cite{3,4}, which
cannot, in principle, be explained in conventional quark-models, have
found own comfortable places in the corresponding KK towers.

Our predictions concerning the masses of hadron states are model
independent, they are related with the fundamental hypothesis on
existence of the extra dimensions with a compact internal extra space
only. In general, each storey in the KK towers is degenerated, i.e.
it may contain several flats for hadron states with the different
quantum numbers but with approximately equal masses. In addition, the
hadron states with the same quantum numbers may have different masses
depending on what KK tower they live in, or in other words depending
on what decay mode the hadron states have been observed in. This
difference in the masses might be measured in the experiments with a
high mass resolution. We have already discussed this non-trivial fact
in analysing the SELEX measurements; see details in \cite{10}.

It should be especially emphasized that KK excitation corresponding
to a certain storey in the given KK tower may have exactly the same
quantum numbers which have been extracted from partial wave analysis
because the definite KK tower corresponds to the definite decay
channel what the given partial wave analysis has been done for.

As mentioned above all KK excitations are very narrow, they have the
widths about a few MeV; see Eq. (\ref{width}). The broad peaks in
hadron spectra may appear in our approach as an envelope of the
narrow KK excitations predicted by the Kaluza-Klein scenario. We have
an idea that an availability of non-diagonal elements in the KK
towers might play the crucial role in understanding the broad peaks
in hadron spectra. This idea has to be explored in the nearest
future.

It is well known that the pole positions extracted form partial wave
analyses have the least model dependence compared to other parameters
such as widths, (in)elasticity, couplings and so on. Our predictions
for the masses of KK excitations are strong, that is wy we have
performed at the moment only the comparison of the calculated mass
values for the KK excitations with the masses of the $N$ and $\Delta$
baryons determined from the partial wave analyses.

In conclusion, we would like once again to claim that all well
established $N$ and $\Delta$ baryons are excellently incorporated in
the created unified picture for hadron spectra. No doubt, new
experiments with a higher mass resolution and sensitivity would be
very helpful to obtain new, more accurate and more full data of high
quality. In that case it would be possible to make a reavaluation of
all known data to refit the total baryon spectrum and to remove the
existing discrepancies. We expect that such new experiments will be
set up in the near future for this goal.

\newpage
\vspace*{-1cm}
\begin{center}
Table 1. Kaluza-Klein tower of KK excitations for the $p\pi$ system
and $N$ baryons.

\vspace{5mm}
\begin{tabular}{|c|c|c|l|l|}\hline\vspace{-4.5mm}
& & & & \\
 n & $ M_n^{p\pi^\pm}$MeV & $ M_n^{p\pi^0}$MeV & N BARYONS (PDG) [St:\{$\star$\}] &\hspace{12mm} $ M_{exp}^{N\pi}$\,MeV  \\ \hline
1  &  1084.79 & 1080.39  &  &    1080.6, 1086.5 MAMI \\
2  &  1104.30 & 1100.37  &  &    1104, 1113 SPES4 \\
3  &  1133.48 & 1130.08  &  &    1130.4(8), 1133.7 SPES3 \\
4  &  1169.65 & 1166.72  &  &    1166.9, 1170.4 SPES3 \\
5  &  1210.92 & 1208.38  &  &    1198, 1202 SPES4\\
6  &  1256.07 & 1253.85  &  &    1251.1 SPES3 \\
\hline
\end{tabular}

\end{center}

\newpage
\vspace*{-1cm}

\begin{center}
Table 2. Kaluza-Klein tower of KK excitations for the $n\pi$ system
and $N$ baryons.

\vspace{5mm}
\begin{tabular}{|c|c|c|l|l|}\hline\vspace{-4.5mm}
& & & & \\
 n & $ M_n^{n\pi^0}$MeV & $ M_n^{n\pi^\pm}$MeV & N BARYONS (PDG) [St:\{$\star$\}] &\hspace{12mm} $ M_{exp}^{N\pi}$\,MeV  \\
 \hline
1  &  1081.69  & 1086.08  &  &    1080.6, 1086.5 MAMI \\
2  &  1101.66  & 1105.59  &  &    1104, 1113 SPES4 \\
3  &  1131.36  & 1134.76  &  &    1130.4(8), 1133.7 SPES3 \\
4  &  1168.00  & 1170.92  &  &    1166.9, 1170.4 SPES3 \\
5  &  1209.64  & 1212.18  &  &    1198, 1202 SPES4 \\
6  &  1255.10  & 1257.32  &  &    1251.1 SPES3 \\
\hline
\end{tabular}
\end{center}

\newpage
\vspace*{-1cm}

\begin{center}
Table 3. Kaluza-Klein tower of KK excitations for the $p\pi$ system
and $\Delta$ baryons.

\vspace{5mm}
\begin{tabular}{|c|c|c|l|l|} \hline \vspace{-4.5mm}
& & & & \\
n & $ M_n^{p\pi^\pm}$MeV & $ M_n^{p\pi^0}$MeV & $\Delta$ BARYONS
(PDG) [St:\{$\star$\}]&\hspace{12mm} $ M_{exp}^{N\pi}$\,MeV \\ \hline
1  &  1084.79 & 1080.39  &  &  \\
2  &  1104.30 & 1100.37  &  &  \\
3  &  1133.48 & 1130.08  &  &  \\
4  &  1169.65 & 1166.72  &  &  \\
5  &  1210.92 & 1208.38  &  &  \\
6  &  1256.07 & 1253.85  &  &  \\
\hline
\end{tabular}
\end{center}

\newpage
\vspace*{-1cm}

\begin{center}
Table 4. Kaluza-Klein tower of KK excitations for the $n\pi$ system
and $\Delta$ baryons.

\vspace{5mm}
\begin{tabular}{|c|c|c|l|l|}\hline\vspace{-4.5mm}
& & & & \\
n & $ M_n^{n\pi^0}$MeV & $ M_n^{n\pi^\pm}$MeV & $\Delta$ BARYONS (PDG) [St:\{$\star$\}] & $\hspace{12mm} M_{exp}^{N\pi}$\,MeV  \\
 \hline
1  &  1081.69  & 1086.08  &  &  \\
2  &  1101.66  & 1105.59  &  &  \\
3  &  1131.36  & 1134.76  &  &  \\
4  &  1168.00  & 1170.92  &  &  \\
5  &  1209.64  & 1212.18  &  &  \\
6  &  1255.10  & 1257.32  &  &  \\
\hline
\end{tabular}
\end{center}

\newpage
\vspace*{1cm}

\begin{center}
Table 5. KK excitations for the $N\pi$ system and narrow exotic low
mass baryons \cite{3,4}.

\vspace{5mm}
\begin{tabular}{|c|c|c|c|c|l|}\hline \vspace{-4.5mm}
& & & & &  \\
n & $ M_n^{p\pi^\pm}$MeV & $ M_n^{p\pi^0}$MeV & $ M_n^{n\pi^0}$MeV &
$ M_n^{n\pi^\pm}$MeV & $ \hspace{14mm} M_{exp}^{N\pi}$\,MeV  \\
\hline
1  &  1084.79 & 1080.39  & 1081.69 & 1086.08 & 1080.6, 1086.5 MAMI     \\
2  &  1104.30 & 1100.37  & 1101.66 & 1105.59 & 1104, 1113 SPES4        \\
3  &  1133.48 & 1130.08  & 1131.36 & 1134.76 & 1130.4(8), 1133.7 SPES3 \\
4  &  1169.65 & 1166.72  & 1168.00 & 1170.92 & 1166.9, 1170.4 SPES3    \\
5  &  1210.92 & 1208.38  & 1209.64 & 1212.18 & 1198, 1202 SPES4        \\
6  &  1256.07 & 1253.85  & 1255.10 & 1257.32 & 1251.1 SPES3            \\
\hline
\end{tabular}
\end{center}

\newpage

\begin{center}
Table 6. Kaluza-Klein tower of KK excitations for the $p\pi$ system
compared with the masses of the  $N$ baryons determined from the KSU
\cite{13}, KH \cite{12} and CMB \cite{11} analyses.

\vspace{5mm}
\begin{tabular}{|c|c|c|c|l|l|l|}\hline\vspace{-4.5mm}
& & & & \\
 n & $ M_n^{p\pi^\pm}$MeV & $ M_n^{p\pi^0}$MeV & N(PDG) [St:\{$\star$\}] & \hspace{5mm}KSU &\hspace{5mm} KH & \hspace{4mm}CMB \\ \hline
1  &  1084.79 & 1080.39  &  &   & & \\
2  &  1104.30 & 1100.37  &  &   & & \\
3  &  1133.48 & 1130.08  &  &   & & \\
4  &  1169.65 & 1166.72  &  &   & & \\
5  &  1210.92 & 1208.38  &  &   & & \\
6  &  1256.07 & 1253.85  &  &   & & \\
\hline
\end{tabular}
\end{center}

\newpage

\begin{center}
Table 7. Kaluza-Klein tower of KK excitations for the $p\pi$ system
compared with the masses of the $\Delta$ baryons determined from the
KSU \cite{13}, KH \cite{12} and CMB \cite{11} analyses.

\vspace{5mm}
\begin{tabular}{|c|c|c|c|l|l|l|} \hline \vspace{-4.5mm}
& & & & \\
n & $ M_n^{p\pi^\pm}$MeV & $ M_n^{p\pi^0}$MeV & $\Delta$(PDG)
[St:\{$\star$\}] & \hspace{5mm}KSU &\hspace{4mm} KH & \hspace{4mm}CMB
\\ \hline
1  &  1084.79 & 1080.39  &  &  & &  \\
2  &  1104.30 & 1100.37  &  &  & &  \\
3  &  1133.48 & 1130.08  &  &  & &  \\
4  &  1169.65 & 1166.72  &  &  & &  \\
5  &  1210.92 & 1208.38  &  &  & &  \\
6  &  1256.07 & 1253.85  &  &  & & \\
\hline
\end{tabular}
\end{center}

\newpage

\begin{center}
Table 8. Kaluza-Klein tower of KK excitations for the $p\pi\pi$
system and $\Delta$ baryons.

\vspace{5mm}
\begin{tabular}{|c|c|c|c|l|l|}\hline\vspace{-4.5mm}
& & & & & \\
 n & $ M_n^{p\,2\pi^0}$MeV & $ M_n^{p\,\pi^0\pi^\pm}$MeV & $
 M_n^{p\,2\pi^\pm}$MeV & $\Delta$ BARYON & \hspace{12mm} $M_{exp}^{\Delta}$\,MeV  \\
 \hline
1  & 1221.60 & 1226.00 & 1230.40  & $\Delta[1232]$  $P_{33}$ & 1231$\pm$1 Manley\hfill 92 \\
2  & 1258.80 & 1262.73 & 1266.66  &  & \\
3  & 1313.67 & 1317.07 & 1320.47  &  & \\
4  & 1380.61 & 1383.54 & 1386.47  &  & \\
5  & 1455.84 & 1458.38 & 1460.91  &  & \\
6  & 1536.98 & 1539.20 & 1541.42  & $\Delta[1600]$  $P_{33}$  & 1522$\pm$13 Hoehler \hfill 79 \\
\hline
\end{tabular}
\end{center}

\newpage

\begin{center}
Table 9. Kaluza-Klein tower of KK excitations for the $n\pi\pi$
system and $\Delta$ baryons.

\vspace{5mm}
\begin{tabular}{|c|c|c|c|l|l|}\hline\vspace{-4.5mm}
& & & & & \\
n & $ M_n^{n 2\pi^0}$MeV & $ M_n^{n\,\pi^0\pi^\pm}$MeV & $
 M_n^{n 2\pi^\pm}$MeV & $\Delta$ BARYON & \hspace{12mm}$ M_{exp}^{\Delta}$\,MeV  \\
 \hline
1  & 1222.89 & 1227.29 & 1231.69  & $\Delta[1232]$  $P_{33}$ & 1231$\pm$1 Manley\hfill 92 \\
2  & 1260.09 & 1264.02 & 1267.95  &  & \\
3  & 1314.95 & 1318.35 & 1321.75  &  & \\
4  & 1381.89 & 1384.82 & 1387.74  &  & \\
5  & 1457.10 & 1459.64 & 1462.17  &  & \\
6  & 1538.23 & 1540.45 & 1542.67  & $\Delta[1600]$  $P_{33}$  & 1522$\pm$13 Hoehler \hfill 79 \\
\hline
\end{tabular}
\end{center}

\newpage

\begin{center}
\vspace*{-15mm}

Table 10. Kaluza-Klein tower of KK excitations for the $N\rho$ system
and $N,\Delta$ baryons.

\vspace*{5mm}
\begin{tabular}{|c|c|c|l|l|}\hline
 n & $ M_n^{n\rho}$ MeV & $ M_n^{p\,\rho}$ MeV   & \hspace{5mm} $ M_{exp}^{\,N}$\,MeV &\hspace{4mm} $ M_{exp}^{\,\Delta}$\,MeV \\ \hline
1  & 1710.90 & 1709.61 & N[1710]$P_{11}$         & $D_{33}$(1710$\pm$30)  \\
2  & 1716.98 & 1715.69 & $P_{11}$(1717$\pm$28)   &                        \\
3  & 1727.07 & 1725.79 & N[1720]$P_{13}$         &                        \\
4  & 1741.09 & 1739.82 &                         &
$P_{31}$(1744$\pm$36)  \\ \hline
5  & 1758.95 & 1757.69 & $P_{11}$(1766$\pm$34)   & $\Delta[1750]P_{31}$   \\
   &         &         &                         & $F_{35}$(1752$\pm$32)  \\ \hline
6  & 1780.53 & 1779.28 &                         &                        \\
\hline
\end{tabular}
\end{center}

\newpage

\begin{center}
\vspace*{-5mm}

Table 11. Kaluza-Klein tower of KK excitations for the $N\eta$ system
and $N,\Delta$ baryons.

\vspace*{5mm}
\begin{tabular}{|c|c|c|l|l|}\hline
 n & $ M_n^{n\eta}$ MeV & $ M_n^{p\eta}$ MeV & \hspace{5mm} $ M_{exp}^{\,N}$\,MeV &\hspace{4mm} $ M_{exp}^{\,\Delta}$\,MeV \\ \hline
1  & 1489.35 & 1488.06 &                          &                          \\
2  & 1496.77 & 1495.48 &                          &                          \\
3  & 1509.04 & 1507.76 &                          &                          \\
4  & 1526.00 & 1524.73 & N[1520]$D_{13}$          &                          \\
5  & 1547.47 & 1546.20 & N[1535]$S_{11}$          &                          \\
6  & 1573.20 & 1571.95 &                          &                          \\
7  & 1602.97 & 1601.73 &                          & $\Delta[1600]P_{33}$     \\
\hline
\end{tabular}
\end{center}

\newpage

\begin{center}
\vspace*{-15mm}

Table 12. Kaluza-Klein tower of KK excitations for the $N\omega$
system and $N,\Delta$ baryons.

\vspace*{5mm}
\begin{tabular}{|c|c|c|l|l|}\hline
 n & $ M_n^{n\omega}$ MeV & $ M_n^{p\omega}$ MeV & \hspace{5mm} $ M_{exp}^{\,N}$\,MeV &\hspace{4mm} $ M_{exp}^{\,\Delta}$\,MeV \\ \hline
1  & 1724.15 & 1722.86 & N[1720]$P_{13}$        & $D_{33}$(1710$\pm$30)  \\
2  & 1730.18 & 1728.89 &                        &                        \\
3  & 1740.17 & 1738.89 &                        &
$P_{31}$(1744$\pm$36)  \\ \hline
4  & 1754.07 & 1752.80 &                        & $\Delta[1750]P_{31}$   \\
   &         &         &                        & $F_{35}$(1752$\pm$32)  \\ \hline
5  & 1771.77 & 1770.51 & $P_{11}$(1766$\pm$34)  &                        \\
6  & 1793.17 & 1791.92 &                        &                        \\
\hline
\end{tabular}
\end{center}

\newpage

\begin{center}
\vspace*{-5mm}

Table 13. Kaluza-Klein tower of KK excitations for the $\Lambda K$
system and $N,\Delta$ baryons.

\vspace*{5mm}
\begin{tabular}{|c|c|c|l|l|}\hline
 n & $M_n^{\Lambda K^\pm}$MeV & $M_n^{\Lambda K^0}$MeV & \hspace{5mm} $ M_{exp}^{\,N}$\,MeV &\hspace{4mm} $ M_{exp}^{\,\Delta}$\,MeV \\ \hline
1  & 1611.87 & 1615.85 &                          & $\Delta[1600]P_{33}$     \\
2  & 1619.36 & 1623.30 &                          & $\Delta[1620]S_{31}$     \\
3  & 1631.72 & 1635.60 &                          &                          \\
4  & 1648.77 & 1652.56 & N[1650]$S_{11}$          &
\\ \hline
5  & 1670.27 & 1673.96 & N[1675]$D_{15}$          &                          \\
   &         &         & N[1680]$F_{15}$          &                          \\ \hline
6  & 1695.97 & 1699.54 & N[1700]$D_{13}$          &
$\Delta[1700]D_{33}$     \\
\hline
\end{tabular}
\end{center}

\newpage

\begin{center}
\vspace*{-5mm}

Table 14. KK excitations for the $(\Sigma K)^\pm$ system and
$N,\Delta$ baryons.

\vspace*{5mm}
\begin{tabular}{|c|c|c|c|l|l|}\hline
 n & $ M_n^{\Sigma^+K^0}$MeV & $ M_n^{\Sigma^-K^0}$MeV & $
 M_n^{\Sigma^0K^\pm}$MeV &\hspace{3mm} $ M_{exp}^{\,N}$\,MeV &\hspace{6mm} $ M_{exp}^{\,\Delta}$\,MeV \\
 \hline
1  & 1689.49 & 1697.59 & 1688.78  & N[1680]$F_{15}$        &                          \\
2  & 1696.80 & 1704.88 & 1696.12  & N[1700]$D_{13}$        & $\Delta[1700]D_{33}$     \\
3  & 1708.86 & 1716.91 & 1708.24  & N[1710]$P_{11}$        &                          \\
4  & 1725.49 & 1733.51 & 1724.94  & N[1720]$P_{13}$        &
\\ \hline
5  & 1746.48 & 1754.46 & 1746.02  &                        & $\Delta[1750]P_{31}$     \\
   &         &         &          &                        & $F_{35}$(1752$\pm$32)    \\ \hline
6  & 1771.57 & 1779.50 & 1771.20  & $P_{11}$(1766$\pm$34)  &                          \\
\hline
\end{tabular}
\end{center}

\newpage

\begin{center}
Table 15. KK excitations for the $(\Sigma K)^{0,++,--}$ system and
$N,\Delta$ baryons.

\vspace*{5mm}
\begin{tabular}{|c|c|c|c|l|l|}\hline
 n & $ M_n^{\Sigma^+K^\pm}$MeV & $ M_n^{\Sigma^-K^\pm}$MeV & $
 M_n^{\Sigma^0K^0}$MeV &\hspace{3mm} $ M_{exp}^{\,N}$\,MeV &\hspace{6mm} $ M_{exp}^{\,\Delta}$\,MeV \\
 \hline
1  & 1685.51 & 1693.60 & 1692.76  & N[1680]$F_{15}$        &                          \\
2  & 1692.86 & 1700.94 & 1700.06  & N[1700]$D_{13}$        & $\Delta[1700]D_{33}$     \\
3  & 1704.98 & 1713.04 & 1712.11  & N[1710]$P_{11}$        &                          \\
4  & 1721.70 & 1729.73 & 1728.73  & N[1720]$P_{13}$        &
\\ \hline
5  & 1742.79 & 1750.78 & 1749.70  &                        & $\Delta[1750]P_{31}$     \\
   &         &         &          &                        & $F_{35}$(1752$\pm$32)    \\ \hline
6  & 1768.00 & 1775.93 & 1774.77  & $P_{11}$(1766$\pm$34)  &                          \\
\hline
\end{tabular}
\end{center}

\end{document}